\documentclass[prl,twocolumn,floatfix,showpacs,amsmath,amssymb]{revtex4}
\usepackage{bbm}
\usepackage{amsfonts}
\usepackage{graphicx}
\usepackage{color}
\usepackage{amsmath}
\usepackage{amssymb}
\usepackage{latexsym}
\usepackage{psfrag}

\begin{document}

\preprint{Tarasov et al.}

\title{Quantized magnetic confinement in quantum wires}

\author{A. Tarasov} \author{S. Hugger} \altaffiliation[Present address: ]{Fraunhofer Institut f\"ur Angewandte
Festk\"orperphysik, Tullastr. 72, 79108 Freiburg, Germany.}\author{Hengyi Xu}\author{M. Cerchez}\author{T. Heinzel}
\email{thomas.heinzel@uni-duesseldorf.de}
\affiliation{Heinrich-Heine-Universit\"at, Universit\"atsstr. 1,
40225 D\"usseldorf, Germany}
\author{I. V. Zozoulenko}
\affiliation{Solid State Electronics, Department of Science and Technology, Link\"{o}ping
University, 60174 Norrk\"{o}ping, Sweden}
\author{U. Gasser-Szerer}
\affiliation{Solid State Physics Laboratory, ETH Z\"urich, CH-8093 Z\"urich, Switzerland}
\author{D. Reuter}\author{A. D. Wieck}
\affiliation{Lehrstuhl f\"ur Angewandte Festk\"orperphysik, Ruhr-Universit\"at Bochum, 44780 Bochum, Germany
}
\date{\today}

\begin{abstract}
Ballistic quantum wires are exposed to longitudinal profiles of perpendicular magnetic fields composed of a spike (\emph{magnetic barrier}) and a homogeneous part. An asymmetric magnetoconductance peak as a function of the homogeneous magnetic field is found, comprising quantized conductance steps in the interval where the homogeneous magnetic field and the magnetic barrier have identical polarities, and a characteristic shoulder with several resonances in the interval of opposite polarities. The observations are  interpreted in terms of inhomogeneous diamagnetic shifts of the quantum wire modes leading to magnetic confinement.
\end{abstract}

\maketitle

Low-dimensional electron gases exposed to inhomogeneous magnetic fields have revealed a variety of fascinating transport features in recent years, for example magnetic superlattices \cite{Ye1995}, snake and cycloid orbits, \cite{Novoselov2002,Nogaret2003} or rectification \cite{Hara2004a}. Due to its elementary character, the \emph{magnetic barrier} (MB), i.e. a magnetic field oriented perpendicular to a two-dimensional electron gas (2DEG) which is strongly localized in the transport (\textit{x})-direction and homogeneous in the transverse (\textit{y})-direction, has been at the focus of  recent transport experiments \cite{Leadbeater1995,Monzon1997,Johnson1997,Kubrak2000,Bae2007,Hong2007,Vancura2000,Cerchez2007,Hugger2007} and a good qualitative \cite{Leadbeater1995,Monzon1997,Johnson1997,Kubrak2000,Bae2007,Hong2007} as well as quantitative \cite{Vancura2000,Cerchez2007,Hugger2007} understanding has been developed, based on a solid foundation of theoretical work.\cite{Peeters1993,Matulis1994,Ibrahim1997,Governale2000,Guo2002,Zhai2002,Zhai2005,Majumdar1996,Xu2001,Lu2002,Jiang2002,Seo2004,Jalil2005,Zhai2006,Xu2007a}
Recent theory has been focusing on the quantum properties of MBs, which are expected to be particularly striking in ballistic quantum wires (QWRs). Such systems have been suggested as tunable spin filters \cite{Majumdar1996,Xu2001,Lu2002,Jiang2002,Seo2004,Jalil2005,Zhai2006} and should show transmission resonances as a function of the MB strength or the Fermi energy, \cite{Governale2000,Zhai2005,Zhai2006,Xu2007a} Interest in magnetic barriers has been furthermore intensified lately by their potential to confine Dirac fermions in graphene.\cite{Martino2007} This large body of theoretical work on MBs in QWRs, however,  has experienced surprisingly few experimental tests.\cite{Hugger2008}\\
Here, we report the experimental observation of magnetically bound electronic states in a QWR. Transport measurements on a QWR containing a MB are performed as a function of a superimposed, homogeneous perpendicular magnetic field $B_h^z$ which displaces the MB along the $B^z$ axis. The results are  interpreted in terms of  $x$ - dependent diamagnetic shifts of the wire modes which create confined states when the MB and $B_h^z$ are of opposite polarity.\\
\begin{figure}[ht]
\includegraphics[scale=1.0]{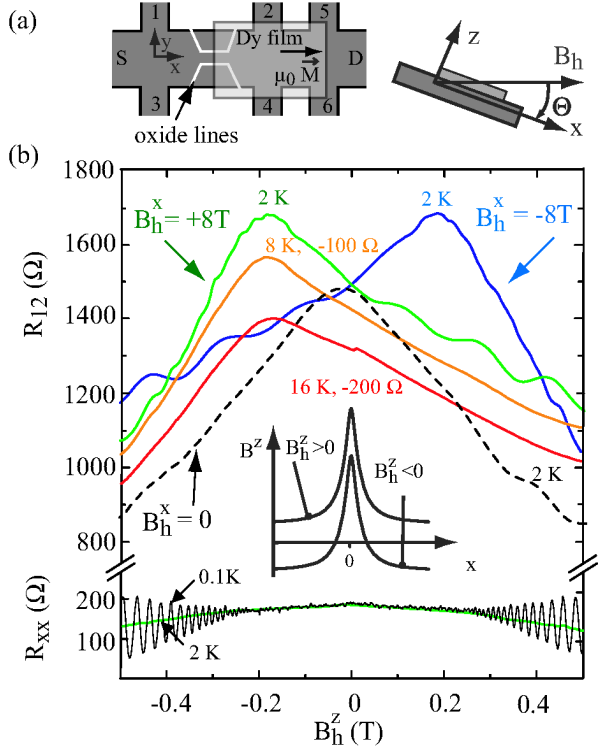}
\caption{(color online). (a) Scheme of the sample layout (left). The QWR is defined by oxide lines (white lines) and partly covered by a Dy film. Voltage probes are labeled as 1 to 6, while S and D denote the source and drain contacts. Right: definition of the tilt angle with respect to the coordinate system and the homogeneous magnetic field $B_h$. (b) Resistance of the QWR system as a function of $B_h^z$ for $B_h^x=\pm 8\,\mathrm{T}$  and $0\,\mathrm{T}$ at various temperatures. Adjacent traces for different temperatures are offset by $100\,\mathrm{\Omega}$ for clarity. The lower part shows the longitudinal resistance of the pristine 2DEG in homogeneous perpendicular magnetic fields, measured over equal area to that one between contacts 1 and 2. Inset: sketch of $B^z(x)$ for different values of $B_h^z$.}
\label{AT2Fig1}
\end{figure}
A $\mathrm{GaAs/Al_xGa_{1-x}As}$ heterostructure with a 2DEG residing $45\,\mathrm{nm}$ below the surface (electron density $3.8\times 10^{15}\,\mathrm{m^{-2}}$, mobility $42\,\mathrm{m^2V^{-1}s^{-1}}$ at a temperature of $2\,\rm{K}$) is used.  The sample layout is depicted in Fig. \ref{AT2Fig1}(a). A Hall bar with Ohmic contacts is prepared by optical lithography. The QWRs are defined by scanning probe lithography. \cite{Held1999}  Their lithographic length is $\approx 500\,\mathrm{nm}$ and their width  is varied in different samples from $400\,\mathrm{nm}$ to $600\,\mathrm{nm}$. Subsequently, the structure is covered by a  Cr layer of $10\,\mathrm{nm}$  thickness, and one edge of a ferromagnetic film made from Dy (thickness  $t=250\,\mathrm{nm}$) was aligned close to the center of the QWR, oriented along the $y$-direction by electron beam lithography and metallization. The opposite Dy edge is located  inside a Hall cross for magnetometery.\cite{Monzon1997} Measurements at temperatures $T\geq 2\,\mathrm{K}$ were performed in a $\mathrm{^4He}$ gas flow cryostat, while for $T<1\,\mathrm{K}$, a $\mathrm{^3He/^4He}$ dilution refrigerator was used. Both systems are equipped with a superconductive magnet and a rotatable sample stage, such that the orientation of the homogeneous magnetic field $B_h$ could be adjusted between parallel to the QWR ($x$-direction in Fig. \ref{AT2Fig1}(a)) and perpendicular to the 2DEG ($z$-direction). Parallel orientation with an accuracy of  $\pm 0.05^{\circ}$ is checked by a vanishing Hall voltage between contacts 1 and 3. In this magnetic field, the Hall resistances of the 2DEG and of the Hall cross containing the MB differ by an offset of $40\,\mathrm{\Omega}$ for tilt angles $\Theta \leq 4^{\circ}$, corresponding to $B_h^z=570\,\mathrm{mT}$, indicating that the Dy magnetization stays in-plane and remains saturated.\cite{Hugger2007} We use the standard expression for the MB shape \cite{Vancura2000}
\begin{equation}
B_{MB}^z (x)=-(\mu_0 M)/(4\pi)\ln
[(x^2+z^2_0)/(x^2+(z_0+t)^2] \label{eq1}
\end{equation}
where $\mu_0M$ denotes the magnetization of the Dy film, while  $z_0=55\,\mathrm{nm}$ is the distance of the 2DEG to the Dy layer. At our maximum Dy magnetization of  $1.9\,\mathrm{T}$,  eq. \eqref{eq1} gives a MB amplitude of $B_{peak}=520\,\mathrm{mT}$ and a full width at half maximum of $360\,\mathrm{nm}$. The polarities of $B_h$ and $\Theta$ determine the magnetic field profile $B^z (x,\Theta)=B_{MB}^z(x)+B_h^z(\Theta)$ (inset in Fig. \ref{AT2Fig1}(b)). In a field of $B_h=8\,\mathrm{T}$, $R_{12}$ is measured as a function of $\Theta$. We have investigated three samples of this kind with the above parameters in five cooldowns, all showing qualitatively similar behavior. Here, we focus on one device (a QWR of $400\,\mathrm{nm}$ width and  $500\,\mathrm{nm}$ length) and on data collected during one cooldown in each cryostat.\\
\begin{figure}[ht]
\includegraphics[scale=1.0]{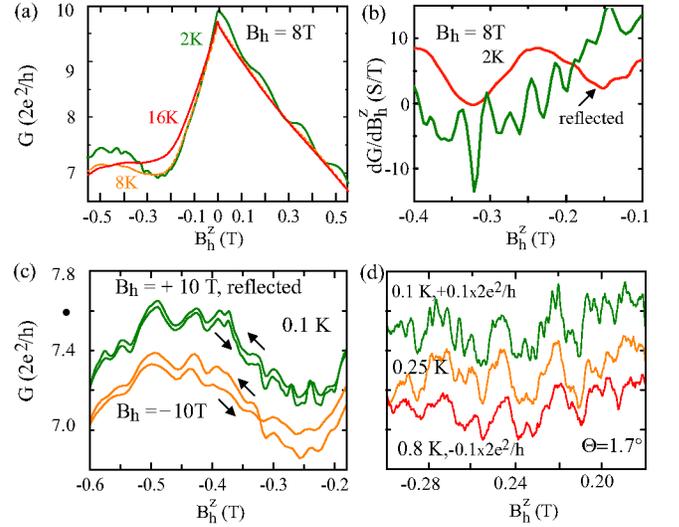}
\caption{(color online). (a) Conductance of the QWR obtained from the transformation of the data in Fig.1. (b)  The derivative of the trace at $2\mathrm{K}$ in the range of the resonances, in comparison to the data in the range of the steps for  $B_h^z>0$, reflected about $B_h^z=0$. (c) Close-up of the conductance resonances at $0.1\,\mathrm{K}$ for positive and negative polarity of the MB. The trace for $B_h=+10\,\mathrm{T}$ is reflected about $B_h^z=0$, and the arrows denote the sweep directions. (d) Temperature dependence of the conductance resonances as a function of $B_h^z$, tuned by sweeping $B_h$ at a fixed tilt angle of $\Theta =1.7^{\circ}$. The traces are offset for clarity.}
\label{AT2Fig2}
\end{figure}
Fig. \ref{AT2Fig1}(b) shows  $R_{12}$ as a function of $B_h^z$ for  different $B_h^x$. For $B_h^x =+8\,\mathrm{T}$, an asymmetric peak with a maximum at $B_h^z=-200\,\mathrm{mT}$ is observed. The slope of its left flank is roughly twice the slope of its right flank. Resistance steps are visible for $B_h^z \geq 0$ while the steep flank contains some weakly pronounced, reproducible features. This magnetoresistance gets reflected about $B_h^z=0$ as the magnetization of the Dy film is reversed ($B_h =-8\,\mathrm{T}$). None of these features can  be attributed to Shubnikov-de Haas oscillations which are represented in the lower part of Fig. \ref{AT2Fig1}(b). The overall peak shape depends only weakly on the temperature and is still present at $16\,\mathrm{K}$, while the resistance steps and the resonances are barely visible at $T= 8\,\mathrm{K}$. For comparison we also show $R_{12}(B_h^z$) for $B_h^x = 0$, where we observe the well-known symmetric, negative magnetoresistance.\cite{Houten1988}\\
\begin{figure}[ht]
\includegraphics[scale=1.0]{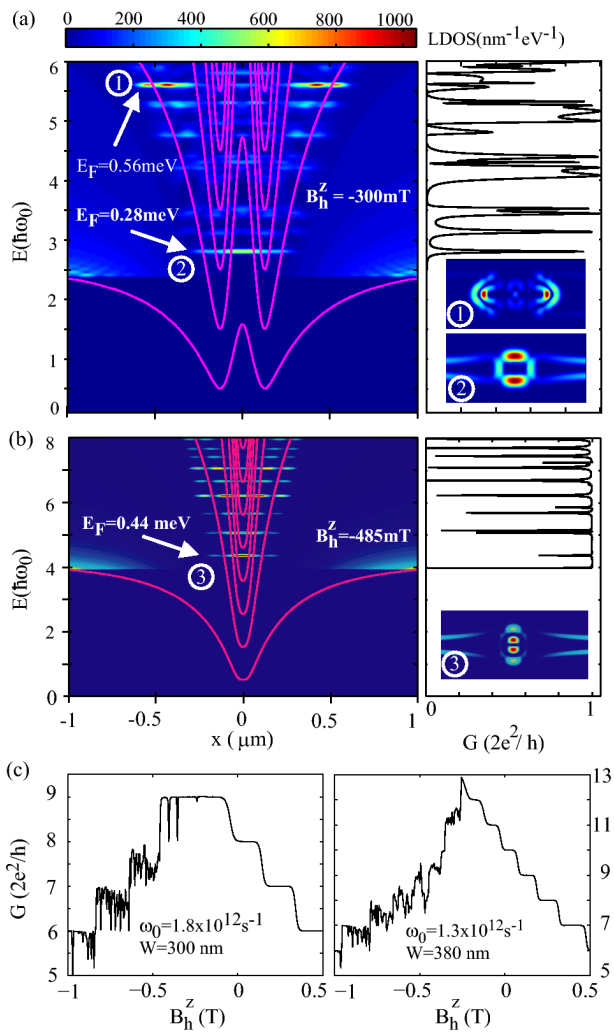}
\caption{(color online). (a) Left: mode energies (red lines) and the LDOS integrated along the $y$ - direction, as a  function of the energy and $x$ for a MB with $B_{peak}=+470\,\mathrm{mT}$ and $B_h=-300\,\mathrm{mT}$. The minima of the modes are located at the $x$ values where $B^z$ vanishes. The right part shows the corresponding conductance. The LDOS as a function of $x$ and $y$ for the states at $E=280\,\mathrm{\mu eV}$ (bottom) and $E=560\,\mathrm{\mu eV}$ (top) is shown in the inset. (b) Left part: mode energies and the LDOS as in (a)  for $B_h=-480\,\mathrm{mT}$, and the conductance (right).  The inset shows the LDOS in the $xy$ plane for the state at $E=440\,\mathrm{\mu eV}$. (c) Calculated conductance of parabolic QWRs  with two confinement strengths $\omega_0$ exposed to a magnetic barrier as above, as a function of the $B_h^z$. $W$ denotes the QWR width at $E_F= 12\,\mathrm{meV}$.}
\label{AT2Fig3}
\end{figure}
For a discussion of the data and their comparison to simulations, we transform them into the two-terminal conductance $G$ of the QWR. According to the Landauer-B\"uttiker formalism,\cite{Landauer1957,Buttiker1988} the transformation reads \cite{Haug1988,Houten1988} $G=1/ (R_{12}-R_{b}+|R_{13}|)$. Here, $R_b= 110\,\mathrm{\Omega}$ denotes the background longitudinal resistance between contacts 1 and 2.  $G$ as determined from the measurements, shown in Fig. \ref{AT2Fig2}(a), has a maximum at $B_h^z=0$ while the maximum of $R_{12}$ transforms into a minimum of $G$ at $B_h^z\approx -250\,\mathrm{mT}$. The conductance plateaus for $B_h^z>0$ are at $\approx j \times 2e^2/h$ with $j$ being an integer. The conductance fluctuations at $B_h^z\lesssim -0.1\,\mathrm{T}$ have a maximum amplitude of $\approx 0.15\times 2e^2/h$. The slope of $G(B_h^z)$ in this range varies by roughly the same amount as in the range of the conductance steps for $B_h^z>0$, but the period is a factor of $\approx 7$ smaller, Fig. \ref{AT2Fig2} (b). These features are characterized further by measurements below $1\,\mathrm{K}$ in the dilution refrigerator. In  Fig. \ref{AT2Fig2} (c),  their reproducibility and their invariance with respect to inversion of $B^z(x)$ is shown: reversing the polarity of $B_h^x$ (and with it the MB) as well as the tilt angle corresponds to a reflection of $G$ about $B_h^z=0$. Here, the heat generated by rotating the sample smears the structure to some extent. We therefore fix the tilt angle at $1.7^{\circ}$ in a complementary experiment and sweep $B_h$ from $-10\,\mathrm{T}$ corresponding to $B_h^z=0.3\,\mathrm{T}$ to $-6\,\mathrm{T}$ ($B_h^z=-0.18\,\mathrm{T}$), Fig. \ref{AT2Fig2} (d). Under these conditions, the magnetization of the Dy film remains almost constant and oriented along the $x$ - direction (the MB remains fixed), but the diamagnetic shift induced by $B_h^x$ is varied in addition to $B_h^z$ such that the traces are not directly comparable to those in (c). Similar fluctuations showing more fine structure due to the lower effective electron temperatures are observed, with a weak temperature dependence between $0.1\,\mathrm{K}$ and $0.8\,\mathrm{K}$.\\
We proceed by interpreting the magnetoconductance with guidance from numerical simulations based on a combination of the tight-binding model with recursive Green's functions at zero temperature. The formalism is described in detail elsewhere.\cite{Xu2007a,Hugger2008} We first look at the basic properties of a model system in the limit of few occupied modes. For this purpose, we choose a QWR with a parabolic confinement given by $\omega_0=1.52\times 10^{11}\,\mathrm{s^{-1}}$ corresponding to a mode spacing of  $0.1\,\mathrm{meV}$, and  a magnetic barrier of a shape according to eq. \eqref{eq1} with $B_{peak}=0.47\,\mathrm{T}$. With $B_h^z$ superimposed, the local density of states (LDOS) as well as the conductance are calculated as a function of the Fermi energy in Fig. \ref{AT2Fig3}. The wire modes have an $x$-dependent energy of $E_j(x)=\hbar\omega (x)(j+1/2)$ with $\omega(x)=\sqrt{\omega_0^2+\omega_c^2(x)}$, where $\omega_c(x)=eB^z(x)/m^*$ is the local cyclotron frequency, and $j$ is an integer. This profile shows a single or a double quantum well structure for $B_{peak}<-B_h^z$ and for $B_{peak}>-B_h^z>0$, respectively, in which bound states are formed. The scenario for $B_h^z > 0$ is similar to $B_h^z = 0$ which has already been discussed theoretically in Ref. \onlinecite{Xu2007a}. Fig. \ref{AT2Fig3}(a) shows the results for the double well scenario, $B_h^z=-300\,\mathrm{mT}$.  The energies and spatial extensions of the bound states are visualized by integrating the LDOS along the $y$-direction. Due to the presence of several modes, a large number of bound states exists even at low mode occupation numbers. The LDOS at the energies of two bound states, i.e. at $E=280\,\mathrm{\mu eV}$ and $560\,\mathrm{\mu eV}$ are shown as a function of $x$ and $y$ in the insets. Two types can be distinguished, resembling bonding and antibonding states. The state at $E=280\,\mathrm{\mu eV}$ is ring-like, which can be interpreted in a semiclassical picture as a combination of snake orbits in $y$-direction along the two lines of $B^z=0$ and edge states along the $x$-direction. The second type of state is exemplified by the LDOS at $E=560\,\mathrm{\mu eV}$, which is concentrated at the potential minima but very small at the wire edges. For Fermi energies below $E_j(x=\pm\infty)$, $G$ is zero. As soon as the first mode is occupied throughout the system, transmission resonances are observed (right part of Fig. \ref{AT2Fig3} (a)) at the energies of the bound states. A systematic inspection indicates that the bonding ring-like states generate lead to transmission maxima, while at the energies of antibonding states, regions of very small LDOS across the QWR are always observed (insets in Fig. \ref{AT2Fig3} and reflection resonances are found.\\
For $-B_h^z>B_{peak}>0$, the bound states are all dot-like and the character of the conductance changes, see Fig. \ref{AT2Fig3}(b). As the first mode gets occupied, the conductance jumps from zero to $2e^2/h$ and gets then suppressed at the energies of the confined states, indicating resonant reflection. These states are similar in their transport character to the antibonding states of the double well. Again, the LDOS at these energies in the $x-y$ - plane (insets) is suppressed all across the QWR close to the bound state. Common to all resonances is that they originate from  the interference of  propagating states with magneto-electrically bound states belonging to higher modes, giving them a Fano character.\\
From these elementary studies, it becomes apparent that in our experiment with up to 10 occupied modes, many resonances of both characters are to be expected. In order to relate the model to the experiment, we simulate the conductance with realistic parameters. The Fermi energy is set to $12\,\mathrm{meV}$, and a harmonic confinement of $\omega_0=1.3\times 10^{12}\,\mathrm{s^{-1}}$  and $\omega_0=1.8\times 10^{12}\,\mathrm{s^{-1}}$ is assumed, leading to wire widths of $380\,\mathrm{nm}$ and $300\,\mathrm{nm}$, respectively. The conductance is calculated as a function of $B_h^z$ for a fixed magnetic barrier of the shape and amplitude as described above. The experimentally observed structure is reproduced qualitatively, see Fig. \ref{AT2Fig3}(c). For $B_h^z \geq 0$ the conductance drops in quantized steps, and no resonances are found. This reflects magnetic depopulation which is strongest at the MB peak position. For $B_h^z < 0$, the overall conductance drops as well due to the diamagnetic shift outside the magnetic barrier, with sharp transmission resonances of both characters seen in Fig. \ref{AT2Fig3}(a) and (b) superimposed. The simulations also suggest that the shoulder has its origin in a quantized conductance step which is modified by the superposition of many overlapping resonances. The shape of the magnetoconductance around $B_h^z =0$ depends sensitively on the parameters chosen, in particular on the energy of the uppermost occupied mode with respect to $E_F$.  In the simulations, the conductance maximum is always found at $B_h^z<0$, while in the experiment, the maximum is at $B_h^z \approx 0$. Most likely, the reason for this deviation is that in the experiment, the wire is shorter than in the simulations. For small values of $B_h^z$, the minima of the double potential formed by the QWR modes lie outside the QWR and the simulations no longer represent the experimental situation. However, we cannot include the widening of the QWR into a 2DEG into the model in its present form, while measuring in longer wires would drive the system out of the ballistic regime.\cite{Hugger2008} Finally, the simulated structure is  richer than the measured one, which we attribute to thermal smearing not considered  in the simulations.\\
In conclusion, we have demonstrated magnetic confinement in quantum wires by displacing a magnetic barrier in B-space, which can generate single-and double-well potentials. Further experiments suggest themselves in order to investigate spin and interaction effects in this system. Moreover, implementing the presented concept in graphene to achieve magnetic confinement \cite{Martino2007} may prove possible.\\
The authors thank K. Ensslin (ETH Z\"urich) for allocating measurement time in his dilution refrigerator for this project, as well as assistance by J. G\"uttinger.  Financial support from  the Heinrich-Hertz foundation (T.H.), the Forschungs-F\"orderungsfonds of the HHU (M.C.) and from BMBF nanoQuit as well as SFB 491 (D.R. and A.D.W.) is acknowledged.

\end{document}